\documentclass[9pt]{Interspeech2024}

\usepackage{xcolor}
\usepackage{booktabs}
\usepackage{siunitx}

\interspeechcameraready

\title{USM RNN-T model weights binarization}

\usepackage{listings}

\usepackage{algorithm}
\usepackage{algpseudocode}

\usepackage{amsmath,graphicx}
\usepackage{color}
\usepackage{setspace}
\definecolor{codegreen}{rgb}{0,0.6,0}
\definecolor{codegray}{rgb}{0.5,0.5,0.5}
\definecolor{codepurple}{rgb}{0.58,0,0.82}
\definecolor{backcolour}{rgb}{1.00,1.00,1.00}
\lstdefinestyle{mystyle}{
    backgroundcolor=\color{backcolour},   
    commentstyle=\color{codegreen},
    keywordstyle=\color{magenta},
    numberstyle=\tiny\color{codegray},
    stringstyle=\color{codepurple},
    basicstyle=\ttfamily\footnotesize,
    breakatwhitespace=false,         
    breaklines=true,                 
    captionpos=b,                    
    keepspaces=true,                 
    numbers=left,                    
    numbersep=5pt,                  
    showspaces=false,                
    showstringspaces=false,
    showtabs=false,                  
    tabsize=2
}
\name{Oleg}{Rybakov}
\name{Dmitriy}{Serdyuk}
\name{Chengjian}{Zheng}
\address{Google LLC, USA}

\email{\{rybakov,dserdyuk,cjzheng\}@google.com}


\begin{document}

\maketitle

\begin{abstract}
    
    Large-scale universal speech models (USM) are already used in production. However, as the model size grows, the serving cost grows too. Serving cost of large models is dominated by model size that is why model size reduction is an important research topic. In this work we are focused on model size reduction using weights only quantization. We present the weights binarization of USM Recurrent Neural Network Transducer (RNN-T) and show that its model size can be reduced by 15.9x times at cost of word error rate (WER) increase by only 1.9\% in comparison to the float32 model. It makes it attractive for practical applications.
    
\end{abstract}

\section{Introduction}

In the last several years the size of automatic speech recognition~(ASR) models increased by more than 10x: from hundreds of millions \cite{li2019jasper, gulati2020conformer} to several billions parameters~\cite{whisper, pratap2023scaling, usm}. Serving cost goes up with model size increase, that is why ASR models compression is a hot research topic.
Sparse network pruning~\cite{takeda17_pruning, Prune_Quant_EDGE, lai2021parp, emili2023neural, jiang23d_interspeech} as well as quantization~ \cite{fasoli20214bit, bie2020simplified, asr_4bit, asr_2bit, usm_2bit} are successfully applied on end-to-end ASR. In \cite{asr_4bit, asr_2bit} authors show that it is possible to quantize 160M parameters model (trained on ~900 hours of Librispeech data~\cite{Librispeech}) with 4 bits and 2 bits with no accuracy loss. But at the same time they show that a 160M parameters model quantized on a production data set (with millions of hours of speech) has 1.4\% (absolute) WER degradation for 2 bits quantization \cite{asr_2bit}. Accuracy reduction can be even higher, e.g. in \cite{asr1bit2bit4bit, usm_2bit} for 2 bits weights only quantization WER is increased by several times. To address this issue, authors in \cite{usm_2bit} explored the combination of quantization with sparsity. Based on these observation we can conclude that 4 bits weights quantization can be quality neutral on most of ASR models, 2 bits quantization works on some set of models (it depends on both model and data size and can require more experiments) and as expected 1 bit quantization (aka binarization) is the most challenging (e.g. authors of \cite{asr1bit2bit4bit} reported ~3x accuracy degradation), so it motivates us to investigate quantization aware weights binarization.

Models binarization is popular topic, e.g. it is explored for computer vision \cite{courbariaux2016binarized, rastegari2016xnornet, he2023bivit, bulat2019xnornet}, language~\cite{NEURIPS2022_5c1863f7, wang2023bitnet, ma2024era} and speech recognition~\cite{xiang17_interspeech, Qian2019BinaryNN, asr1bit2bit4bit} applications. Standard weights binarization~\cite{courbariaux2016binarized,rastegari2016xnornet, xiang17_interspeech} is based on sign function to binarize weights to \{-1, 1\}. It uses Straight-Through Estimator to overcome non-differentiability of the sign function. There can be stochastic or deterministic sign function~\cite{courbariaux2016binarized}. It can be combined with batch normalization~\cite{courbariaux2016binarized} and \emph{hard tanh}~\cite{xiang17_interspeech} to deal with the gradient. The output of the binarized weight can be scaled by \textit{absmean}~\cite{rastegari2016xnornet, wang2023bitnet}. As shown in~\cite{rastegari2016xnornet} \textit{absmean} is an optimal scaler for binarization problem. Group Quantization~\cite{wang2023bitnet} or sub-channel quantization~\cite{asr_2bit} can be applied to parallelize weights quantization (and it also can reduce quantization error). In this work we propose a combination of sub-channel quantization (also called block-wise~\cite{subChannel}) with binarization based on \textit{absmean}~\cite{rastegari2016xnornet, wang2023bitnet, NEURIPS2022_5c1863f7}. As in~\cite{NEURIPS2022_5c1863f7, wang2023bitnet} we use \textit{absmean} binarization but without weights centralization. It simplifies model quantization and does not impact model accuracy in comparison to quantization with weights centralization.

Our main contributions are outlined as below:
\begin{itemize}[leftmargin=*]
    \item We simplify USM-CTC model training by switching to USM-RNNT model and reducing the number of training procedures from 4 to 2 with no accuracy loss (in comparison to baseline USM-CTC model). We show that with a reduced number of training procedures USM-RNNT is more friendly for quantization. With 2bits weights quantization USM-RNNT has only 1.4\% WER increase in comparison to 3x WER degradation in the USM-CTC model.
    \item We develop a new approach of USM-RNNT model binarization. It combines sub-channel quantization with \textit{absmean} binarization. It is open sourced at~\cite{praxis}.
    \item We benchmark proposed binarization approach on USM RNN-T 1 Billion model trained on millions of hours of speech data and show that  model size can be reduced by 15.9x times at cost of 1.9\% (absolute) WER increase in comparison to the float32 baseline model.
\end{itemize}

The rest of the paper is organized as follows. In Section~\ref{sec:model-data}, we describe the baseline model, the RNN-T model we use for quantization, and the datasets. In Section~\ref{sec:quantization}, we present our quantization approaches. We report and discuss our experimental results in Section~\ref{sec:experiments}. Finally, we review related work in Section~\ref{sec:related} and conclude in Section~\ref{sec:conclusion}.

\section{Model and Data}
\label{sec:model-data}

\subsection{CTC-based Universal Speech Model}
\label{sec:usm-ctc}
We base our work on Google Universal Speech Model~(USM~\cite{usm}). This is a single large model which performs ASR on a multitude of languages. This model uses a conformer encoder~\cite{gulati2020conformer} and a CTC loss~\cite{graves2006connectionist}. The USM uses a complex training procedure, therefore below we outline only the most important components.

The USM-CTC model consists of a conformer encoder which requires the 128-dimensional filter-bank features as input.
A trainable language embedding is injected as a first token to the encoder. This allows the model to work with about $300$ languages. The encoder features are fed into the CTC decoder.

The model is trained in a 4-step procedure:
\begin{enumerate}
    \item First, a 600M parameter CTC model is trained on a small supervised set. This model is used as a teacher for the Noisy Student Training~\cite{nst} procedure. Both supervised and unsupervised sets are relabeled with this model to produce pseudo labels.
    \item Then, the encoder is pre-trained with BEST-RQ~\cite{bestrq}.
    \item Then, the full model is pre-trained with the supervised relabeled subset obtained in Step 1.
    \item Finally, the model is fine-tuned with task-specific data.
\end{enumerate}

\subsection{RNN-T-based Universal Speech Model}
\label{sec:usm-rnnt}
We expand the USM-CTC model and simplify its training.

We use the RNN-T loss~\cite{Graves2012-ad,rnnt} which is known to have higher performance. In particular, we use a linear joint network and a 3-layer LSTM decoder. A downside of using the RNN-T instead of CTC is that the training may be less stable. We address this issue by using the Adafactor~\cite{adafactor} optimizer. Furthermore, Adafactor allowed us to reduce the memory footprint of the model during training.

Using a stronger loss allowed us to exclude the BEST-RQ pre-training and the fine-tuning steps. Therefore, our approach has two steps:
\begin{enumerate}
    \item Similarly to the USM-CTC, we use a 600M CTC model trained on a supervised set to relabel supervised and unsupervised speech datasets.
    \item Then we simply train the RNN-T USM on the mix of all the supervised and semi-supervised data  obtained in Step 1.
\end{enumerate}

In this paper we use a USM RNN-T model with 893 million parameters, we call it the USM RNN-T 1B model. It has a good ratio of performance to number of parameters. The USM RNN-T model is composed of a mel spectrogram followed by a feature extractor followed by 16 conformer layers with a decoder. Mel spectrogram returns frames with 128 features (where every frame is generated per every 40ms with 20ms step). The feature extractor is composed of a convolution layer with kernel size (3, 3) channel size (128, 32) and stride (2, 2), followed by a normalization layer with ReLU activation function. These layers are repeated two times so that the input signal is subsampled by 4x in time dimension. After that output features are projected to 1536 dimensions and it is augmented with positional embedding. Output of the feature extractor is processed by sequence of conformer layers.
Conformer layer parameters are: the model hidden size is equal to the input size (1536); the kernel size of convolution is 5; local self attention uses left and right context equal 129 and 128 accordingly; number of heads in the attention is 12. Conformer layer is open sourced at~\cite{praxis}.  Output of the conformer layers is processed by a standard RNN-T decoder, which has a joint network and predictor. Where the predictor has 3 LSTM layers with 640 hidden units. The joint network also has 640 units. Decoder vocabulary size is 16,384.

In Table~\ref{tab:rnnt-wer} we verify that our model is comparable to the previously published results. We conclude RNN-T simplified training produces results comparable to the baseline USM-CTC model. 

\begin{table}[t]
    \centering
    \caption{\textit{Comparison of non-quantized models on the YouTube test set (WER, \%). Note that the USM-CTC results are not strictly comparable to ours. }
    }
    \begin{tabular}{lrr}
        \toprule
        Method & en\_us & multi-lang \\
        \midrule
        \hspace{3pt} \emph{Published Baselines} \\
        USM-CTC & 13.7 & 26.7 \\
        \midrule
        \hspace{3pt} \emph{Our Results} \\
        USM-RNNT 1B & 8.8 & 25.3 \\
        USM-RNNT 2B & 8.3 & 24.9 \\
        \bottomrule
    \end{tabular}
    \label{tab:rnnt-wer}
\end{table}

\subsection{Data}
\label{sec:data}

We used two datasets for training and one dataset for testing:
\begin{enumerate}

    \item \textbf{Semi-supervised:} this is a large dataset extracted from YouTube videos. We extracted segments of lengths 30 seconds. We ensured that the extracted segments contain speech with high probability. This dataset contains videos of varying quality and a mix of formal and informal speech. The set contains 75 languages and its total length is about 4.2 million hours. We produce pseudo-labels for this set described in the procedure described in Section~\ref{sec:usm-ctc}
    \item \textbf{Supervised:} is a smaller dataset also extracted from YouTube videos. Similarly to the semi-supervised set, we extracted the segments and ensured that they contain speech. In contrast to the semi-supervised set, we used a heuristic approach to include only high quality videos. This dataset was labeled by the human transcribers. Then, we used it to train a teacher model (see Section~\ref{sec:usm-ctc}). Finally, we produced the pseudo-labels labels for this set. We call the result a \textbf{supervised-relabeled} set. This set contains 56 languages and its total length is about 680,000 hours.
    \item \textbf{YouTube test set:} the test set contains a combination of various topics for each given language. The human transcribers labeled this dataset.
\end{enumerate}

\section{Quantization approaches}
\label{sec:quantization}
We apply quantization-aware training on USM RNNT 1B model by quantizing weights of linear and projection attention layers in the Conformer encoder, during training of this model. Note that we keep convolution layers and 3 decoder RNN layers in float format because their size is negligible in comparison to the size of the encoder.

\subsection{2bits quantization}
\label{sec:2bit_quant}
Several methods of 2 bits weights quantization are presented in~\cite{asr_2bit} (they are based on \textit{absmax} quantization). Here we explore two methods from~\cite{asr_2bit}:
\begin{enumerate}
\item  Asymmetric per channel quantization. It uses scale backpropagation~\cite{qiu2023rand}. In \cite{asr_2bit} it is labeled as \textit{I2WasymSc}, here we tagged it as \textit{Exp1}
\item  Asymmetric per channel quantization with scale backpropagation, clipping and sub channel split. We use subchannel quantization with block size equal 64. In \cite{asr_2bit} it is labeled as \textit{I2WasymScSubchClip}. Here we labeled it as \textit{Exp2}.
\end{enumerate}
Please refer to~\cite{asr_2bit} for more details about the above approaches.

\subsection{1bit quantization}
\label{sec:1bit_quant}
We explore several methods of model binarization.
The first one is based on \textit{absmax} asymmetric quantization, labeled as \textit{I2WasymSc} in \cite{asr_2bit}. In this work we set the number of bits equal to 1 and apply static clipping, we label it as \textit{Exp3}.

\begin{figure}[t]
    \centering
    \includegraphics[width=1.0\columnwidth]{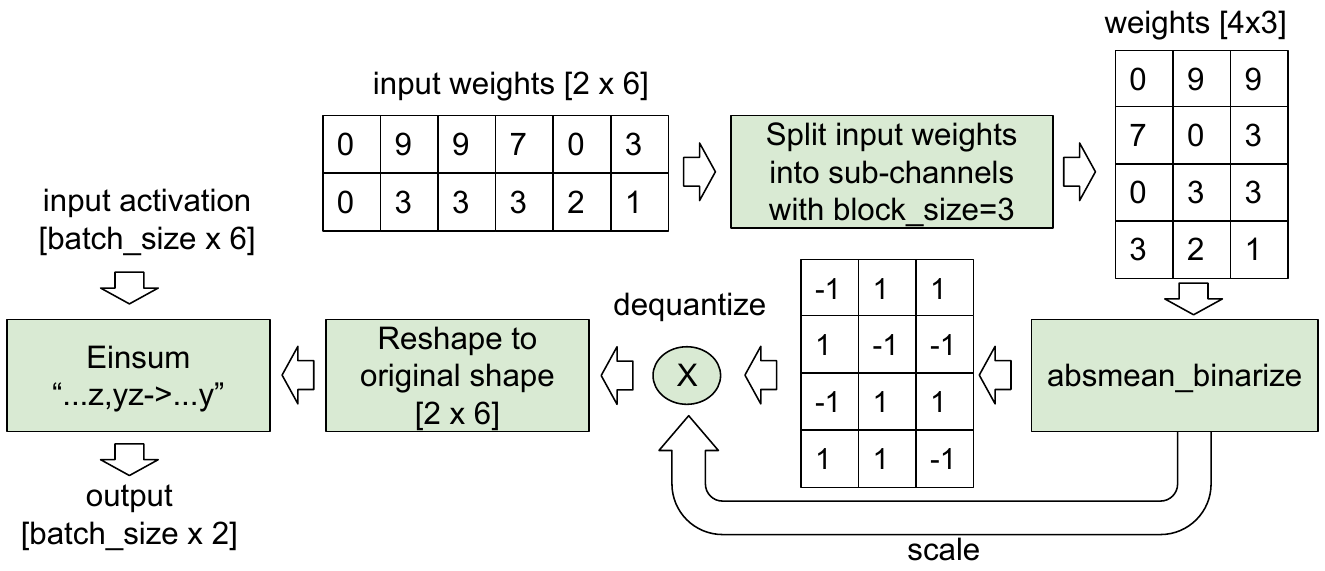}
    \caption{Combination of sub-channel split with \textit{absmean} binarization during training.}
    \label{fig:per_channel}
\end{figure}

Another method of binarization is based on \textit{absmean}~\cite{wang2023bitnet, NEURIPS2022_5c1863f7}. It is shown on Figure~\ref{fig:quant_functions}: input weights \textit{x} can be centralized by subtracting per channel mean value (in line 10 on Figure~\ref{fig:quant_functions} we disable it). Then weights \textit{x} are binarized (in line 15) and the dequantization scale is estimated as per channel \textit{absmean} (in line 14), that is why it is called \textit{absmean} binarization. Note that we make all zeros equal to small epsilon number (in line 11), so that \textit{sign} function returns only 1 and -1 (as shown on Figure~\ref{fig:per_channel} output of \textit{absmean\_binarize} has only 1 and -1 values). To deal with the gradient of a \textit{sign} function we use straight-through estimator based on function \textit{ste}, shown in line 18 in Figure~\ref{fig:quant_functions}. Note that we do not apply weights centralization as in \cite{NEURIPS2022_5c1863f7, wang2023bitnet}, because there was no accuracy difference and it simplifies model quantization. We labeled this approach as \textit{Exp4}.

We combine \textit{Exp4} binarization technique with sub channel quantization~\cite{asr_2bit, dettmers2023qlora}, which has a predefined block size equal 64. We labeled this approach as \textit{Exp5}. It is explained in  Figure~\ref{fig:per_channel}, where input weights have shape [2x6]. We increase the number of channels in the input weights by splitting weights into sub-channels with block size equal 3 (it is selected for illustration purpose). After that weight shape becomes [4x3]. These weights are quantized using \textit{absmean} binarization approach~\cite{wang2023bitnet, NEURIPS2022_5c1863f7} as shown on Figure~\ref{fig:quant_functions}. We use fake quantization aware training, so binarized weights are dequantized by multiplying them with scale, as shown on Figure~\ref{fig:per_channel}. Then the dequantized weights are reshaped back to the original weight shape: [2x6]. The last step is \textit{einsum}, which is computed over input activation with dequantized weights. Note that during the inference dequantization step (multiplication by \textit{scale}) is done on \textit{einsum} output, also input activation will be reshaped to match the shape of binarized weights (its shape [4x3] as shown on Figure~\ref{fig:per_channel}).

\begin{figure}[t]
  \centering
\begin{lstlisting}[language=python]
def absmean_binarize(
    x, 
    contract_dims, 
    centralize=False
):
  if centralize:
    mean = jnp.mean(x, 
                    axis=contract_dims, 
                    keepdims=True)
    x = x - mean    
  x = jnp.where(x == 0.0, eps, t)
  scale = jnp.mean(jnp.abs(x), 
                   axis=contract_dims, 
                   keepdims=True)
  x = ste(x, jnp.sign)
  return x, scale

def ste(x, fn):
    return x - jax.lax.stop_gradient(x) + 
           jax.lax.stop_gradient(fn(x))
\end{lstlisting}  
  \caption{\textit{Absmean} binarization function.}
  \label{fig:quant_functions}
\end{figure}

\section{Experiments}
\label{sec:experiments}

\subsection{Baseline USM RNN-T 1B model}
\label{sec:float_baseline}

USM RNNT 1 Billion model described in Section~\ref{sec:usm-rnnt} is trained on 256 TPU v4~\cite{jouppi2023tpu} for 1.2 days. It takes 200,000 training iterations  to converge. The model is trained on about 5 millions of hours of speech as described in section~\ref{sec:data}. Its mean WER is equal to 25.3\% (computed over 61 languages) and model size in bits are presented in Table~\ref{tab:model_size}. This experiment is labeled as \textit{Exp0} in Table~\ref{tab:model_size}.

\subsection{USM RNN-T 1B model with 2bits weights quantization}

We run quantization-aware training of \textit{Exp1} and \textit{Exp2} (described in section~\ref{sec:2bit_quant}) on 256 TPU v5 for 270,000 training iterations (it takes 1.7 days).

Results of USM RNN-T 1B with 2bits quantization approaches \textit{Exp1} and \textit{Exp2} are presented in Table~\ref{tab:model_size}. \textit{Exp1} has mean WER 26.7\% and it is only 1.4 points worse than its float baseline. In contrast, USM CTC~\cite{usm_2bit} has 3x WER degradation with 2bits weights quantization. We hypothesize that quantization difficulties of the USM CTC model can be due to multi-stage training procedures, described in section~\ref{sec:usm-ctc} ~\cite{usm_2bit}, and differences in data sets.
The model size of \textit{Exp1} (in bits and estimated model size reduction) is presented in Table~\ref{tab:model_size}. With 2bits weights quantization \textit{Exp1}, we reduce model size by 11.4x times. As expected \textit{Exp2} with sub-channel quantization (block size 64) has a larger quantized model size (due to additional meta data) in comparison to \textit{Exp1}.

\subsection{USM RNN-T 1B model with binarized weights}
Training binarized models \textit{Exp3}, \textit{Exp4} and \textit{Exp5} (described in section~\ref{sec:1bit_quant}) takes 3.4 days (446,000 training iterations) on 256 TPU v5~\cite{V5}. Method \textit{Exp3} diverged and has 100\% WER. But methods \textit{Exp4} and \textit{Exp5} converged and have WER 27.5\% and 27.2\% accordingly. We hypothesize that method \textit{Exp3} diverged because it is based on \textit{absmax} quantization. Its quantization scale coefficient is defined by max value of abs weights and normalization by max value can be less stable in comparison to normalization by mean abs value. Both \textit{Exp4} and \textit{Exp5} are based on \textit{absmean} quantization (scale coefficient is defined by mean value of abs weights). We observe that a model binarized with \textit{Exp5} can reduce model size by 15.9x times at the cost of increasing mean WER by 1.9\% in comparison to float32 baseline.

\begin{figure}[t]
  \centering
  \includegraphics[width=\linewidth]{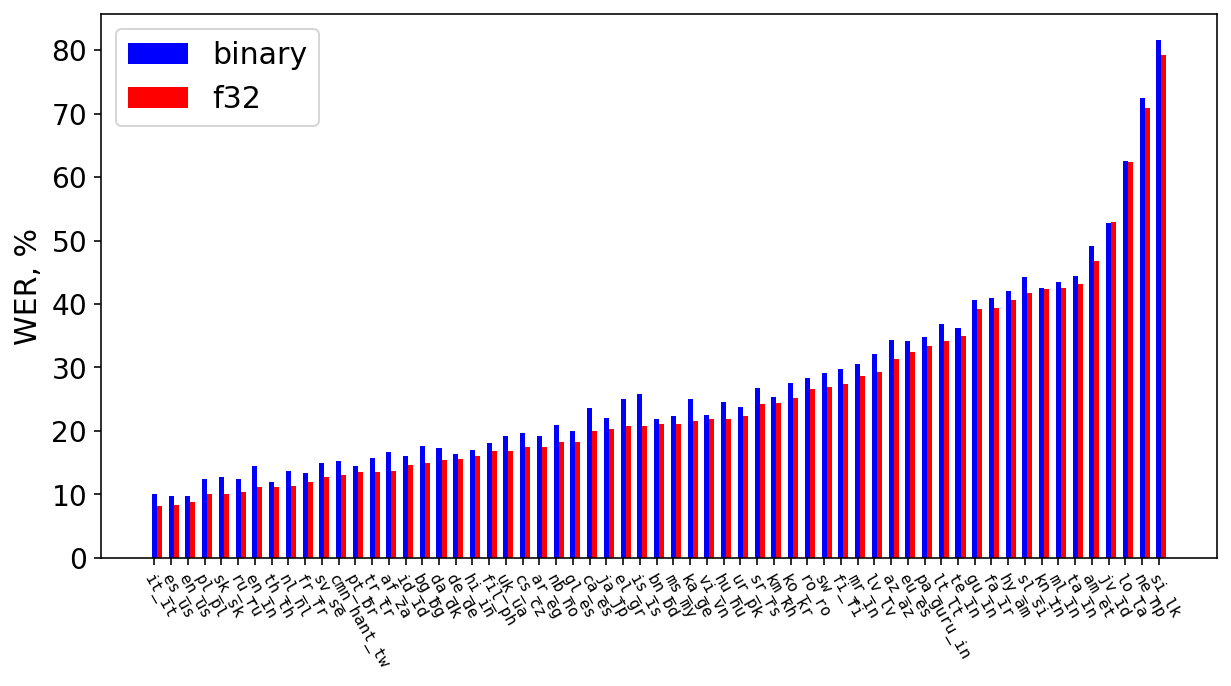}
  \caption{The WER of f32 (Exp0) and binarized (Exp5) models on 61 languages.}
  \label{fig:clean}
  \label{fig:wer_lang}
\end{figure}

The WER comparison of float baseline model \textit{Exp0} vs binarized model \textit{Exp5} over 61 languages is shown on Figure~\ref{fig:wer_lang}. We can see that some languages have WER \textgreater 50\%, we explain it by data quality of such languages.

Limitations of the binarization approach presented in this paper: there still is an accuracy gap between binarized and float baseline models; model binarization can take at least 2x times longer because it needs more time to converge.

\begin{table}[t]
    \centering
    \caption{Results on baseline model and proposed quantization approaches. Mean \textit{WER} over 61 languages, model size[bits] and estimated model \textit{size reduction}, defined as: (float model size) / (quantized model size). For each quantized model, we estimate the quantized size with the quantization metadata (scale only for binary quantization, scale and zero point for 2-bit quantization) stored as either f32 or int8. We estimate the 95\% confidence interval for WER $\pm 0.25$.}
    \begin{tabular}{p{15mm}p{10mm}p{15mm}|p{15mm}}
        \toprule
        Experiment & WER[\%] & \multicolumn{2}{c}{\parbox{3cm}{size[\# bits billion] (size reduction)}}  \\
        \midrule
        \textit{Exp0 (f32)} & 25.3\% & \multicolumn{2}{c}{28.6 (N/A)}  \\
        \midrule
        & \ & \textit{f32 meta} & \textit{int8 meta} \\
        \textit{Exp1(2bit)} & 26.7\% & 2.5 (11.4x) & 2.5 (11.4x) \\
        \textit{Exp2(2bit)} & 27.5\% & 3.6 (7.9x) & 2.8 (10.2x) \\
        \textit{Exp3(1bit)} & N/A & 1.6 (17.9x) & 1.6 (17.9x) \\        
        \textit{Exp4(1bit)} & 27.5\% & 1.6 (17.9x) & 1.6 (17.9x) \\        
        \textit{Exp5(1bit)} & 27.2\% & 2.2 (13x) & 1.8 (15.9x)\\
        \bottomrule
    \end{tabular}
    \label{tab:model_size}
\end{table}

\section{Related work}
\label{sec:related}

In this section we review the prior literature on ASR model quantization, compare and contrast it to our approach.

As in \cite{asr_2bit} we explore quantization of RNN-T model (with 118M parameters) and train it on Librispeech data. We binarize weights of this model using best quantization method (from \cite{asr_2bit}, based on \textit{absmax}) and get only 0.1\% WER increase on Librispeech test clean and 0.3\% WER increase on test other accordingly. Note that we have to train this model two times longer in comparison to float baseline. But as we can see in Table~\ref{tab:model_size}, binarization method \textit{Exp3} based on \textit{absmax}, does not converge on a large model (USM RNN-T, with 1B parameters), trained on a large data set(5 million hours of speech). That is why in this work we are focused on large model binarization trained on large data.

Recent studies are focused on the Universal Speech Model~(USM,~\cite{whisper, chan2021speechstew, Zhang_2022, li2021scaling, li2022asr2k, usm}) which can recognize multiple languages. We base our work on \cite{usm}. These models show state of the art results with large model size (\textgreater 1B parameters). As a result it increases serving cost. A standard approach of speech model serving cost reduction is model compression based on weights quantization~\cite{xiang17_interspeech, Qian2019BinaryNN, asr1bit2bit4bit, fasoli20214bit, bie2020simplified, asr_4bit, asr_2bit, usm_2bit}. In~\cite{usm_2bit} authors quantize weights of USM-CTC model with 2bits but accuracy drops by 3x. So they addressed it by combining quantization with sparsity. In our paper we replace USM CTC by USM RNN-T and simplify training step procedures as a result we can quantize USM RNN-T model (\textit{Exp1} in Table~\ref{tab:model_size}) with 2bits at a cost of 1.4\% WER increase (instead of 3x increase in USM CTC). We explore 2bits quantization approach based on~\cite{asr_2bit}. In addition to 2bits quantization we explore USM RNN-T binarization, which has not been investigated in previous studies. Standard weights binarization is based on \textit{absmean}~\cite{rastegari2016xnornet, wang2023bitnet, NEURIPS2022_5c1863f7}. As in~\cite{NEURIPS2022_5c1863f7, wang2023bitnet} we use \textit{absmean} binarization but without weights centralization. It simplifies model quantization and does not impact model accuracy (on USM model) in comparison to quantization with weights centralization. We combine \textit{absmean} binarization with sub-channel quantization~\cite{asr_2bit} and show that it is possible to binarize USM RNN-T 1B model and reduce its model size by 15.9x times at a cost of 1.9\% WER increase in comparison to baseline float32 model.

\section{Conclusion}
\label{sec:conclusion}
We simplified USM-CTC 1B model model training by replacing it with USM RNNT 1B model and reducing the number of training step procedures. As a result, the quantized USM RNNT 1B model with 2 bits weights quantization has reduced model size by 11.4x times at cost of only 1.4\% (absolute) WER increase vs 3x WER increase of 2bits weights quantization in USM-CTC 1B.
We presented a new weights binarization for the USM RNNT 1B model and showed that model size can be reduced by 13x or 15.9x (if double quantization is applied) at a cost of 1.9\% WER increase in comparison to baseline float32 model. It makes it attractive for real applications.

\section{Acknowledgements}
We would like to thank Jian Li, Eugene Weinstein and Pedro Moreno for their support and valuable discussions.


\bibliographystyle{IEEEtran}
\bibliography{mybib}

\end{document}